\documentclass[useAMS,usenatbib]{mn2e}

\title[Magnetic flux transport]{Type II superconductivity and magnetic flux
transport in neutron stars}
\author[P. B. Jones]{P. B. Jones\thanks{E-mail:
p.jones1@physics.ox.ac.uk}\\
Department of Physics, University of Oxford, Denys Wilkinson Building,
Keble Road, Oxford OX1 3RH\\}
\begin{document}

\date{}

\pagerange{\pageref{firstpage}--\pageref{lastpage}} \pubyear{}

\maketitle

\label{firstpage}

\begin{abstract}
The transition to a type II proton superconductor which is believed
to occur in a cooling neutron star is accompanied by changes in the
equation of hydrostatic equilibrium and by the formation of proton
vortices with quantized magnetic flux.  Analysis of the electron
Boltzmann equation for this system and of the
proton supercurrent distribution formed at the transition
leads to the derivation of a simple expression for the transport
velocity of magnetic
flux in the liquid interior of a neutron star.  This shows
that flux moves easily as a consequence of the interaction between
neutron and proton superfluid vortices during intervals of spin-down
or spin-up in binary systems.
The differences between the present analysis and those of
previous workers are reviewed and an error in the paper
of Jones (1991) is corrected.
\end{abstract}


\section{Introduction}

The origin and modes of evolution of neutron star magnetic fields have
been topics of interest since the early papers of Pacini (1967) and
Gold (1968). We refer to Bhattacharya \& Srinivasan (1995) for a review
of the many later papers published on these problems.
The difficulty in the interpretation of observations on radio pulsars and
binary-system neutron stars has been that the magnetic flux transport
properties of both the solid crust and the liquid interior are not
well known.  Empirical deduction of these properties is not feasible
and {\it a priori} theoretical input is required.  A series of papers
on the structure of the solid phase in the neutron-drip region of the
inner crust have shown that it is amorphous, heterogeneous
in nuclear charge $Z$, with a high temperature-independent resistivity
(Jones 2004).  But there remain uncertainties about the movement
of flux in the liquid interior, the problem with which this paper is
concerned.

The outer region of the liquid core is believed to consist of a
normal system of ultra-relativistic
electrons, a $^{1}S_{0}$ or $^{3}P_{2}$ neutron superfluid, and a
$^{1}S_{0}$ proton superfluid (see Baym \& Pethick 1979).  Negative
$\mu$-mesons, essentially non-relativistic, are also present if the
electron chemical potential exceeds their rest energy.  Our
assumption, following the early paper of Baym, Pethick \& Pines (1969),
is that the protons form a type II superconductor.  Type I
superconductivity will be present in any density region for which
$\kappa = \lambda/\xi < 1/\sqrt{2}$,
where $\lambda$ is the proton penetration
depth and $\xi$ the coherence length, but this condition can be satisfied
only for very small proton energy gaps.  During the early stages of
neutron-star cooling, the proton phase transition, from normal to
superconducting Fermi liquid, is accompanied by the formation of
a mixed state.  On microscopic scales, magnetic flux 
becomes quantized.  The quantum of magnetic flux is
$\phi_{0} = hc/2e = 2.07 \times 10^{-7}$ G cm$^{2}$, and it is
confined to the core of a proton vortex.  
The highest-density regions of the inner crust may consist of
low-dimensional structures (see Pethick \& Ravenhall 1995) in place of
spherical nuclei.  In this case, the boundary of the proton superconductor
is not well-defined but we shall nevertheless assume the existence of
a spherical surface within which macroscopic supercurrents can flow.  The
movement of flux across this surface depends, of course, on the $B(H)$
characteristic of the superconductor and the value of its lower
critical field $H_{c1}$ (see, for example, Fetter \& Hohenberg 1969).

Apart from the effects of ohmic diffusion and Hall drift of the field,
we may assume that an approximation to static hydrodynamic equilibrium
exists at times before
the superconducting phase transition.  But the transition changes this
equilibrium because the components of the proton superconductor stress
tensor are larger than those of the normal-system Maxwell tensor by
factors of the order of $H_{c1}/B$, where $B$ is the mean magnetic
flux density of the superconductor mixed state (Jones 1975,
Easson \& Pethick 1977).  The divergence of the superconductor stress
tensor is a volume force and, with the different
transport properties of the superconducting system, produces a
proton-vortex drift velocity. The force has been referred to as
a buoyancy force and was first estimated by Muslimov
\& Tsygan (1985).  They obtained a drift velocity from its
steady-state equilibrium with a viscous force derived
from the magnetic scattering of electrons by an isolated vortex.
This specific problem was later re-examined by
Harvey, Ruderman \& Shaham (1986).  Further work by Jones (1987, 1991)
and by Harrison (1991) showed that the magnetic viscous
force acting on an isolated vortex moving relative to the electrons
is not relevant to the problem of obtaining a drift velocity and that
it is necessary to consider the interaction of the whole system of
vortices which is arranged as a two-dimensional lattice.
The different ways in which magnetic flux might be expelled, primarily
from normal Fermi systems, were analyzed by Goldreich \& Reisenegger
(1992) who emphasized the importance of the stable stratification
of neutron-star matter and introduced the concept of ambipolar
diffusion, that is, drift of the charged components of a plasma
relative to the neutral such as may occur in the interstellar medium
(see, for example, Spitzer 1968).  The severe constraint imposed
on ambipolar diffusion
by stable stratification was the motivation for the work of
Ruderman, Zhu \& Chen (1998).
  
There are several reasons for writing a further paper on this
problem.  The papers cited above, in so far as they consider type II
superconductivity, are contradictory.  There is also an error in the
paper by Jones (1991).  Given the quantity of observational data on
radio pulsars and on binary systems, there is a possibility
that a clear solution to the problem might reveal interesting
information about the neutron star interior.  The idea that proton
vortices move easily as a result of interaction with neutron vortices
during intervals of spin-down or spin-up is widely assumed in
studies of X-ray binary systems (see Bhattacharya \& Srinivasan 1995)
and unambiguous theoretical confirmation of its validity, or
otherwise, is desirable. 

The approach to the problem made in the present paper uses the fact
that, at the temperature considered here, all superfluid quasiparticles
except those localized in vortex cores have negligibly small number
densities.  Thus it is possible to write down a Boltzmann equation
for the electrons (muons) interacting only with muons (electrons) and
with the proton vortices
and obtain a steady-state solution (Section 2).  The screening of
the lepton current in the superconductor and the proton
force-balance give further equations connecting drift velocities
and chemical potential gradients for leptons and protons (Section 3).
A simple and unambiguous result is obtained for the drift velocity
of the proton vortices (equation 22) and hence for the expulsion of
flux from the interior of the neutron star.  It is the principal
result of this paper, and is shown to be unaffected by possible 
lepton interactions with other degrees of freedom.
In Section 4, we attempt
to analyze the differences between this paper and the work of
each of the sets of authors cited above.  The scope of the paper
is purely technical and it does not consider observational evidence
relevant to magnetic flux evolution.  It excludes the possibility of
type I superconductivity induced by interaction between neutron and
proton superfluids (Buckley, Metlitski \& Zhitnitsky 2004), also
systems of several distinct
superconductors which may be present at very high matter densities
in the inner liquid core.  Unless otherwise stated,
all quantities are defined in a coordinate system
corotating with the solid crust of the star and having angular
velocity ${\bf \Omega}$.

\section{The lepton Boltzmann equation}

The reference system we shall use within the chosen coordinates
is the steady state with
zero vortex drift velocity, $v_{L} = 0$.  A local definition of
electron chemical potential is used, $\mu_{e} = \epsilon_{Fe}$,
where $\epsilon_{Fe}$ is the Fermi energy, including rest
energy.  For present purposes,
this is more convenient than the global definition,
$\mu_{e} = (\epsilon_{Fe}-e\phi)\sqrt{g_{00}}$, in terms of the
time-like component of the metric tensor and the
gravitationally-induced electric potential $\phi$.  This latter
definition was favoured by Harrison
(1991); its equilibrium value $m_{e}c^{2}\sqrt{g_{00s}}$,
defined at the stellar surface, is
constant throughout the star.  In the liquid core,
$\epsilon_{Fe}$ can exceed the muon rest energy so that a muon
chemical potential $\mu_{\mu}$ is also defined. At these energies,
lepton transport can be
considered under the assumption of non-quantizing fields
(see Potekhin 1999).  The classical leptonic
orbits in a plane perpendicular to the proton vortices are
irregular polygons whose vertices represent scattering by
the microscopic magnetic flux density $\tilde{\bf B}$ localized
within the vortex cores.  To be specific, we consider first the
electron component.  Its orbit size can be specified by
the orbit radius in the spatially-averaged magnetic flux density
${\bf B} = \langle\tilde{\bf B}\rangle$.  This is
$r_{B} = \epsilon_{Fe}/eB = 3.3 \times 10^{-7}B_{12}^{-1}$ cm,
for $\epsilon_{Fe} = 100$ MeV, where
$B_{12}$ is the magnetic flux density in units of $10^{12}$ G.
It is several orders of magnitude larger than the intervortex
spacing in the triangular lattice,
$d=4.9\times 10^{-10}B_{12}^{-1/2}$ cm.
For such a system, it is possible to define, within a small
element of phase space, a spatially-dependent electron Fermi
distribution function which satisfies a Boltzmann
equation (see, for example, Pines \& Nozi\`{e}res 1966).
In the steady state, with $v_{L} = 0$ and isotropic
distribution function $n^{0}_{k}$,  this can be
expressed as
\begin{eqnarray}
\lefteqn{{\bf v}_{\bf k}\cdot\nabla n^{0}_{k} -e\left(
\mbox{\boldmath $\mathcal{E}$}_{0}
+ \frac{1}{c}{\bf v}_{\bf k}\times{\bf B} \right)\cdot
\nabla_{\bf k}n^{0}_{k} = }  \nonumber \hspace{2cm}    \\
                &    &  - \Gamma^{0}_{\bf k}
+ e\left(\frac{1}{c}{\bf v}_{\bf k}\times\left(
\tilde{\bf B} - {\bf B}
\right)\right)\cdot\nabla_{\bf k}n^{0}_{k},
\end{eqnarray}
with electron velocity ${\bf v}_{\bf k}$ for momentum ${\bf k}$.
The conservative field $\mbox{\boldmath $\mathcal{E}$}_{0}$  cancels
the $\nabla n^{0}_{k}$ term in the equation.  Motion of the
vortex lattice produces an induction field
$\tilde{\bf E} = - (1/c){\bf v}_{L}\times \tilde{\bf B}$
and changes the electron distribution function to
\begin{eqnarray}
n_{\bf k} = n^{0}_{k} + \beta n^{0}_{k}(1-n^{0}_{k})
\delta\mu_{e} + \delta n_{\bf k},
\end{eqnarray}
in which the first incremental term is isotropic, with
$\beta^{-1} = k_{B}T$, and represents the effect of
a chemical potential change $\delta\mu_{e} = \delta\epsilon_{Fe}
-e\delta\phi$.  The second term is anisotropic.  The modified
Boltzmann equation is then,
\begin{eqnarray}
\lefteqn{{\bf v}_{\bf k}\cdot\nabla n_{\bf k} -e\left(
\mbox{\boldmath $\mathcal{E}$}_{0} + {\bf E}
 + \frac{1}{c}{\bf v}_{\bf k}\times{\bf B}
\right)\cdot\nabla_{\bf k}n_{\bf k} = } \nonumber \hspace{0.5cm}  \\
   &    &  - \Gamma_{\bf k}
+ e\left(\tilde{\bf E} - {\bf E} + \frac{1}{c}{\bf v}_{\bf k}
\times \left(\tilde{\bf B} - {\bf B}\right)\right)\cdot
\nabla_{\bf k}n_{\bf k}
\end{eqnarray}
Both equations have been expressed in terms of the
spatially-averaged fields.  The reason for this is that the
difference between the irregular polygons and the circular
orbits of the spatially-averaged field can be thought of as
the consequence of a scattering
process.  Therefore, the difference terms producing it are
placed on the right-hand side of each equation with the
collision integrals $\Gamma^{0}_{\bf k}$ and
$\Gamma_{\bf k}$.  These are derived from processes other
than magnetic scattering and include electromagnetic scattering
by muons (if present) and by vortex-core quasiparticles.  But these
Boltzmann equations are for a distribution function
$n_{\bf k}({\bf r})$ defined within a small element of
phase space.  Spatial integration over that small element
leaves the terms in them unchanged with the exception of
the difference terms, whose integrals clearly converge to zero
as the linear dimension of the element becomes large compared
with the intervortex spacing $d$.  This serves to establish
the intuitively obvious result (Harrison 1991, Jones 1991)
that, given the condition
$r_{B}\gg d$, the microscopic fields $\tilde{\bf E}$ and
$\tilde{\bf B}$ can be replaced without error by the spatial
averages ${\bf E}$ and ${\bf B}$.

With neglect of some terms of the second order of smallness,
the incremental change in distribution function
$\delta n_{\bf k}$ satisfies
\begin{eqnarray}
-e\delta\mbox{\boldmath $\mathcal{E}$}\cdot
\nabla_{\bf k}n^{0}_{k} -\frac{e}{c}
\left({\bf v}_{\bf k}\times {\bf B}\right)\cdot
\nabla_{\bf k}\delta n_{\bf k} = \Gamma^{0}_{\bf k}
- \Gamma_{\bf k},
\end{eqnarray}
where,
\begin{eqnarray}
\delta\mbox{\boldmath $\mathcal{E}$} =
 {\bf E} + \frac{1}{e}\nabla\delta\mu_{e}.
\end{eqnarray}
In order to write down an explicit form for the collision
integral difference, it is necessary to specify each set of
degrees of freedom with which the electrons interact.
At temperatures well below the superconducting transition,
superfluid quasiparticle
densities are negligible except for those localized in vortex
cores.  The rest frame for these excitations is therefore
that of the vortices, with velocity ${\bf v}_{L}$ in the
reference frame corotating with the solid crust in which
our quantities are defined.  Thus
interaction with them causes
$\delta n_{\bf k}$ to relax to
isotropy in the rest frame of the vortex lattice with relaxation
time $\tau^{e}_{v}$. 
There is also interaction with muons (if present) and possibly with
other excitations, such as zero
sound phonons or other collective modes,
or with  superfluid continuum quasiparticles whose
density will be significant near the transition temperature.
In order to include them, we assume for the muons 
a natural rest frame with velocity ${\bf v}_{\mu}$, in which
their  thermal-equilibrium distribution function is isotropic,
and a separate relaxation time $\tau^{e}_{\mu}$.  For the unspecified
excitations, we assume a rest-frame velocity ${\bf v}_{u}$ and
relaxation time $\tau^{e}_{u}$, with both quantities
treated as unknowns.  Thus
the collision integral difference is
\begin{eqnarray}
\Gamma^{0}_{\bf k} - \Gamma_{\bf k} & = & - \frac{1}{\tau^{e}_{v}}
\left(\delta n_{\bf k} - \beta n^{0}_{k}(1 - n^{0}_{k})
{\bf k}\cdot{\bf v}_{L}\right) \nonumber \\
  &   &  -\frac{1}{\tau^{e}_{\mu}}\left(\delta n_{\bf k} -\beta
  n^{0}_{k}(1-n^{0}_{k}){\bf k}\cdot{\bf v}_{\mu}\right) \nonumber\\
  &   &   - \frac{1}{\tau^{e}_{u}}\left(\delta n_{\bf k}
  - \beta n^{0}_{k}(1 - n^{0}_{k}){\bf k}\cdot{\bf v}_{u}\right).
\end{eqnarray}
Equations (4) and (6) are satisfied by,
\begin{eqnarray}
\delta n_{\bf k} = e\tau^{e}_{v}\frac{\partial n^{0}_{k}}
{\partial\epsilon_{k}}{\bf v}_{\bf k}\cdot{\bf A},
\end{eqnarray}
where ${\bf A}$ is related to the electron current density,
\begin{eqnarray}
{\bf J}^{e} = -N_{e}e{\bf v}_{e} = \frac{-2e}{(2\pi)^{3}}
\int d^{3}k{\bf v}_{\bf k}n_{\bf k} = \sigma^{e}_{v}{\bf A},
\end{eqnarray}
and $N_{e}$ is the electron number density.
Equations (8) define the electron drift velocity ${\bf v}_{e}$.
The resistivities $\left(\sigma^{e}_{j}\right)^{-1}$  are
defined by 
\begin{eqnarray}
\sigma^{e}_{j} = \frac{N_{e}e^{2}\tau^{e}_{j}c^{2}}{\epsilon_{Fe}},
\end{eqnarray}
where the subscript denotes the system with which the electrons
interact. 
The relation between the drift velocities ${\bf v}_{e,\mu,u}$ and
${\bf v}_{L}$ derived from equations (4) and (6) can be
expressed in the form,
\begin{eqnarray}
\lefteqn{\frac{1}{e}\nabla\delta\mu_{e} - \frac{N_{e}e}
{\sigma^{e}_{\mu}}\left({\bf v}_{\mu} - {\bf v}_{e}\right) -
\frac{N_{e}e}{\sigma^{e}_{u}}\left({\bf v}_{u} -
{\bf v}_{e}\right) = }
\nonumber \hspace{1.5cm} \\
   &    &   \frac{N_{e}e}{\sigma^{e}_{v}}
\left({\bf v}_{L} - {\bf v}_{e}\right)
 + \frac{1}{c}
\left({\bf v}_{L} - {\bf v}_{e}\right)\times {\bf B}.
\end{eqnarray}
This unremarkable result is no more than a
force-balance equation relating the force on the
electrons from the vortex lattice (right-hand side) with
the forces exerted by the electrons on the chemical potential
gradient, on the muons, and on the unspecified system of excitations
whose rest-frame has velocity ${\bf v}_{u}$.  A similar equation
exists for the muons,
\begin{eqnarray}
\lefteqn{\frac{1}{e}\nabla\delta\mu_{\mu} - \frac{N_{\mu}e}
{\sigma^{\mu}_{e}}\left({\bf v}_{e} - {\bf v}_{\mu}\right) -
\frac{N_{\mu}e}{\sigma^{\mu}_{u}}\left({\bf v}_{u} -
{\bf v}_{\mu}\right) = }
\nonumber \hspace{1.5cm} \\
   &    &   \frac{N_{\mu}e}{\sigma^{\mu}_{v}}
\left({\bf v}_{L} - {\bf v}_{\mu}\right)
 + \frac{1}{c}
\left({\bf v}_{L} - {\bf v}_{\mu}\right)\times {\bf B}.
\end{eqnarray}
The rate at which the lattice, the muons and the unspecified
excitations do work on unit volume of the electrons
less the rate at which the electron system does work in
moving on the chemical potential gradient is,
\begin{eqnarray}
\lefteqn{N_{e}e{\bf v}_{L}\cdot\left(\frac{N_{e}e}{\sigma^{e}_{v}}
\left({\bf v}_{L} - {\bf v}_{e}\right) + \frac{1}{c}
\left({\bf v}_{L} - {\bf v}_{e}\right)\times {\bf B}\right)}
 \nonumber \hspace{7cm}\\
   - N_{e}{\bf v}_{e}\cdot\nabla\delta\mu_{e}
+ \frac{(N_{e}e)^{2}}{\sigma^{e}_{\mu}}\left({\bf v}_{\mu}
 - {\bf v}_{e}\right)\cdot{\bf v}_{\mu}  \nonumber \\
+ \frac{(N_{e}e)^{2}}{\sigma^{e}_{u}}\left({\bf v}_{u}
 - {\bf v}_{e}\right)\cdot{\bf v}_{u} = \nonumber\hspace{1cm} \\
   \frac{(N_{e}e)^{2}}{\sigma^{e}_{v}}\left({\bf v}_{L}
  - {\bf v}_{e}\right)^{2} + \frac{(N_{e}e)^{2}}{\sigma^{e}_{\mu}}
  \left({\bf v}_{\mu} - {\bf v}_{e}\right)^{2} \nonumber  \\
  +\frac{(N_{e}e)^{2}}{\sigma^{e}_{u}}\left({\bf v}_{u} -
{\bf v}_{e}\right)^{2},
\end{eqnarray}
which is the rate of energy dissipation per unit volume attributable
to ${\bf v}_{e}$.  A similar relation exists for the muons.

The analogous force-balance equation for the proton superfluid
continuum can be written down immediately because the continuum
quasiparticle number density is negligibly small.   It
relates the chemical potential gradient with the Magnus force per
unit volume,
\begin{eqnarray}
N_{p}\nabla\delta\mu_{p} = \frac{N_{p}e}{c}\left({\bf v}_{p0}
 - {\bf v}_{L}\right)\times {\bf B},
\end{eqnarray}
in which ${\bf v}_{p0}$ is the proton drift velocity.
Equations (10), (11) and (13) are the principal results of this
Section but contain a total of 7 variables excluding ${\bf v}_{u}$.
The further
equations required can be obtained by examining the effect of
the normal to superconducting transition on the various
current densities in the crust and liquid interior.

\section{Steady-state equilibrium of the magnetic flux distribution}

Before the transition to superconductivity, the magnetic flux and
lepton current density are slowly-varying functions of
position satisfying Amp\`{e}re's theorem.  If the transition is to
a type II superconductor, it is accompanied on a microscopic scale
by the formation of proton
vortices.  These have quantized magnetic flux
$\mbox{\boldmath $\phi$}_{0}$, locally
collinear with their axes, supported by a circulating
supercurrent distribution whose density decreases exponentially
with radius on the scale of the penetration depth
$\lambda$.  If, on macroscopic scales, the vortex number density
$N_{v}$ is such that the spatially-averaged magnetic flux density
is unchanged, so that $N_{v}\mbox{\boldmath $\phi$}_{0} = {\bf B}$,
the spatial
average of the individual vortex supercurrent distributions
$\tilde{\bf J}^{p\alpha}$ must
satisfy the relation,
\begin{eqnarray}
\left\langle\sum_{\alpha}\tilde{\bf J}^{p\alpha}\right\rangle
 = {\bf J}^{e} + {\bf J}^{\mu}.
 \end{eqnarray}
The pre-existing lepton current density must be screened
out by a supercurrent density ${\bf J}^{p0}$ which, in the
body of the superconductor, has significant variation only
over lengths many orders of magnitude larger than
$\lambda$,
\begin{eqnarray}
{\bf J}^{p0} + {\bf J}^{e} + {\bf J}^{\mu} = 0.
\end{eqnarray}
This ensures that Amp\`{e}re's theorem is
satisfied at all points in space (see Jones 1991) and, as a
consequence of superconductivity, is analogous with the
formation of surface current sheets in the Meissner effect. 
An additional condition expresses the fact that
protons do not cross the superconductor boundary,
\begin{eqnarray}
\left({\bf J}^{p0} + \sum_{\alpha}\tilde{\bf J}^{p\alpha}
\right)_{\perp} = 0.
\end{eqnarray}
The effect of equation (16) is that the supercurrent distribution
${\bf J}^{p0}$ defined by equation (15) must be associated with
a return current sheet at the boundary separating the
superconductor from the normal solid.  
This current sheet, which has negligible kinetic energy,
also maintains the mixed state of the superconductor by
excluding the continuous magnetic flux density of the
normal system.

The neutrality condition $N_{e}+N_{\mu} = N_{p}$ can be regarded
as exact and equation (15) gives the condition
$N_{e}{\bf v}_{e} + N_{\mu}{\bf v}_{\mu} = N_{p}{\bf v}_{p0}$
which is satisfied at all points in the superconductor
by the lepton and proton drift velocities.  The final
equation is for steady-state equilibrium of the
vortex lattice under what is referred to as the buoyancy
force.  The argument for the existence of this quantity is
as follows.  The normal-state hydrostatic equilibrium before
the transition is given by the equation,
\begin{eqnarray}
\rho g_{i} - \frac{\partial P^{0}}{\partial x_{i}}
+ \frac{1}{c}\left(({\bf J}^{e} + {\bf J}^{\mu})
\times{\bf B}\right) = 0,
\end{eqnarray}
in terms of the matter density $\rho$, the gravitational
acceleration ${\bf g}$, and the zero-field pressure $P^{0}$.
The third term is the divergence of the Maxwell stress
tensor.  In hydrostatic equilibrium, its curl is almost
exactly a non-radial vector. (This condition does not apply
in the crust owing to the presence there of a further term,
derived from the solid stress tensor, which allows more
general components in the divergence of the Maxwell
tensor and the resultant phenomenon of Hall drift.  In
the liquid,
$\nabla\times(({\bf J}^{e} + {\bf J}^{\mu})\times {\bf B})$
can produce Hall drift only
of non-radial components of ${\bf B}$.
Here, for brevity, components in the electromagnetic
current density other than ${\bf J}^{e,\mu}$ have been
neglected.)  At the transition, the magnetic part of the
superconductor stress tensor replaces the Maxwell tensor.
Following Easson \& Pethick (1977), we express this in
the form,
\begin{eqnarray}
T^{S}_{ij} = -P^{S}\delta_{ij} + \frac{1}{4\pi}H_{i}B_{j},
\end{eqnarray}
which is a symmetric tensor because ${\bf B}$ and {\bf H}
are locally parallel vectors.
Both isotropic and anisotropic components are larger than
the corresponding Maxwell components by factors of the
order of $H_{c1}/B \gg 1$.  The new hydrostatic equilibrium
is given by
\begin{eqnarray}
\rho g_{i} - \frac{\partial P^{0}}{\partial x_{i}}
 + \frac{\partial T^{S}_{ij}}{\partial x_{j}}
  + f_{Vi} = 0,
\end{eqnarray}
where ${\bf f}_{V}$ is the force per unit volume arising from
interaction between neutron and proton vortices (Sauls 1989),
an additional effect
which is not included in equation (18).  (Strictly, the
existence of ${\bf f}_{V}\neq 0$ depends on rotation of
the neutron superfluid with angular velocity
${\bf \Omega}_{n}\neq{\bf \Omega}$.)
The derivative of the isotropic component of the stress
tensor has been referred to as a buoyancy force, but the
implicit neglect of the anisotropic component means that
this description is not necessarily apt.  Easson \& Pethick
obtained an expression (equation 17 of their paper) for the
isotropic component of the stress tensor valid for any
$\kappa > 1/\sqrt{2}$ in the limit $B \ll H_{c1}$.  But they
note that it can be estimated reliably only in the extreme
type II limit in which $\kappa \gg 1$.  Thus for more general
values of $\kappa$, it is not obvious that the term leads
to a buoyancy force, rather than the reverse.  In the limit
$B \ll H_{c1}$, the magnetic field $H\approx H_{c1}$ and we
can assume that the spatial derivatives of its magnitude
(though not direction) are much
smaller than those of $B$.  With neglect of these terms, and
in the further limit $\kappa \gg 1$, the
divergence of the stress tensor given by equation (18) is,
\begin{eqnarray}
\frac{\partial T^{S}_{ij}}{\partial x_{j}} =
 \frac{1}{4\pi}
\left(\left(\nabla\times{\bf B}\right)\times{\bf H}_{c1}
\right)_{i} + \frac{1}{4\pi}\left(\nabla\cdot{\bf H}_{c1}
\right)B_{i},
\end{eqnarray}
in which ${\bf H}_{c1}$ has magnitude $H_{c1}$ and is
everywhere parallel with ${\bf B}$.
This replaces the third term in equation (19).  The sum of
the third and fourth terms in equation (19) is the total
magnetic flux-dependent force per unit volume 
(defined as ${\bf f}_{B}$) and acts on the vortex
lattice.  Its direction is not necessarily the radial
direction expected of a buoyancy force and it is analogous
with the third term in equation (17).

The lattice drift velocity ${\bf v}_{L}$ is then defined
by equating ${\bf f}_{B}$ with the sum of the force on the
electron and muon
systems and the Magnus force on the superfluid continuum,
\begin{eqnarray}
{\bf f}_{B} - \frac{(N_{e}e)^{2}}{\sigma^{e}_{v}}\left({\bf v}_{L}
 - {\bf v}_{e}\right) - \frac{N_{e}e}{c}\left({\bf v}_{L}
  - {\bf v}_{e}\right)\times{\bf B} \nonumber \hspace{2mm} \\
  - \frac{(N_{\mu}e)^{2}}{\sigma^{\mu}_{v}}\left({\bf v}_{L} -
  {\bf v}_{\mu}\right) - \frac{N_{\mu}e}{c}\left({\bf v}_{L} -
  {\bf v}_{\mu}\right)\times{\bf B} \nonumber \\
  - \frac{N_{p}e}{c}\left({\bf v}_{p0} - {\bf v}_{L}
  \right)\times {\bf B} = 0,
\end{eqnarray}
which gives an extremely simple final result,
\begin{eqnarray}
{\bf v}_{L} = \frac{\sigma^{\mu}_{v}N_{e}^{2}{\bf v}_{e} +
\sigma^{e}_{v}N_{\mu}^{2}{\bf v}_{\mu}}
{\sigma^{\mu}_{v}N_{e}^{2} + \sigma^{e}_{v}N_{\mu}^{2}} +
\frac{\tilde{\sigma}}{(N_{e} + N_{\mu})^{2}e^{2}}{\bf f}_{B},
\end{eqnarray}
where,
\begin{eqnarray}
\tilde{\sigma} = \frac{(N_{e} + N_{\mu})^{2}\sigma^{e}_{v}
\sigma^{\mu}_{v}}{\sigma^{\mu}_{v}N_{e}^{2} + \sigma^{e}_{v}
N_{\mu}^{2}}.
\end{eqnarray}
It is interesting that this result is quite independent of
the existence of the resistivities derived
from lepton interaction with other degrees of freedom
whose natural rest frame is not that of the vortex lattice.
Equations (10), (11), (13), (15) and (21) give ${\bf v}_{L}$ and
a complete description of the system in terms of the
independent variables ${\bf v}_{e}$, ${\bf v}_{\mu}$ and
${\bf f}_{B}$, which all depend on the properties of the normal-state
${\bf B}$-distribution that existed before the superconducting
transition.  Scattering transition rates for $e-p$ and $\mu-p$
at the Fermi surfaces, though of the same order of magnitude, are
not identical so that we anticipate ${\bf v}_{e}\neq {\bf v}_{\mu}$
and retain both as independent variables in equation (22).
Of the remaining variables ${\bf v}_{p0}$ and
$\nabla\delta\mu_{e,\mu,p}$, the increments in chemical
potential,
$\delta\mu_{e}(\delta N_{e}, \delta\phi)$,
$\delta\mu_{\mu}(\delta N_{\mu}, \delta\phi)$ and
$\delta\mu_{p}(\delta N_{p}, -\delta\phi)$, are functions
of just 3 further variables
$\delta N_{e}$, $\delta N_{\mu}$ and $\delta\phi$.
The values of $\tilde{\sigma}$ that are immediately relevant are
those at temperatures below, but within an order of
magnitude of the superconducting
transition temperature.  We refer to Jones (1991) for the
order of magnitude of $\sigma^{e}_{v}$ and for a brief discussion
of the factors determining its temperature dependence.  But
it is possible to assert that as the star cools, it increases
at least as rapidly as $T^{-2}$, where $T$ is the temperature.
Thus at a typical transition temperature of
$3 \times 10^{9}$ K, the
order of magnitude is $\tilde{\sigma} \approx 10^{29} B_{12}^{-1}$
s$^{-1}$.
A force with order of magnitude $f_{B}\approx 10^{20}
B_{12}$ dyne cm$^{-3}$ then leads, from equation (22), to a
velocity $v_{L}\approx 4\times 10^{-7}$ cm s$^{-1}$.
It is evident that $\tilde{\sigma}$, and therefore $v_{L}$
both increase rapidly as the star cools.  Examination of
the orders of magnitude present in equations (5), (10) and (11)
shows that the chemical potential gradients can become quite
large.  But they are almost exactly cancelled by the induction
field.  Our conclusion is that, for any plausible value of
$f_{B}$, the post-transition movement of magnetic flux is
fast in comparison with, for example, radio pulsar spin-down
or binary-system evolutionary time-scales.

\section{Conclusions}

Before proceeding to a comparison of the results obtained
here with those of earlier papers, it is worth considering
two questions about flux movement.  Konenkov \& Geppert (2001)
observed that proton vortices, given a buoyancy
force, always move toward the crust by sliding (with some
dissipation) along the
rectilinear neutron vortices of the rotating neutron superfluid.
(The only exceptional case is that in which proton vortices,
within a small element of solid angle, are approximately
parallel with the neutron vortices.)  This interesting
possibility clearly depends on ${\bf f}_{B}$ having a component,
parallel with the neutron vortices and the spin vector
${\bf \Omega}_{n}$ of the neutron superfluid, with the
appropriate sign.  But we
have seen, both from the discussion given by Easson \& Pethick
and by examination of equation (20), that this sign must depend
on the form of the flux distribution prior to the
superconducting transition.  For example, the force terms
in equation (20) vanish in the case of a uniform ${\bf B}$,
giving ${\bf f}_{B} = {\bf f}_{V}$, which is almost exactly
perpendicular to ${\bf \Omega}_{n}$.
The second question concerns whether
or not proton vortices can be moved inward or outward, relative
to the rotation axis of the star, by interaction with neutron
vortices during intervals of spin-up or spin-down. The neutron
vortices at a distance $r_{\perp}$ from the rotation axis
move with a radial velocity component
$v_{\perp} = -r_{\perp}\dot{\Omega}_{n}/2\Omega_{n}$, where
${\bf \Omega}_{n}$ is the  neutron superfluid angular velocity
at $r_{\perp}$.
From equation (22), ${\bf v}_{L}$  can have a component
of this size provided,
\begin{eqnarray}
\tilde{f}_{V}\tilde{\sigma} > \frac{\pi r_{\perp}\mid\dot{\Omega}\mid
\hbar e^{2}(N_{e}+N_{\mu})^{2}}{4m_{p}\Omega^{2}},
\end{eqnarray}
where $m_{p}$ is the proton mass and $\tilde{f}_{V}$ is the
maximum force that unit length of neutron vortex can exert
on a lattice of proton vortices without intersection.
(Here, we assume ${\bf \Omega}_{n} = {\bf \Omega}$, the
angular velocity of the star.)
It is easy to confirm that this is clearly satisfied
in the case of the observed radio pulsars by
evaluation for the Crab, which gives
$\tilde{f}_{V}\tilde{\sigma} > 8\times 10^{44}$ dyne cm$^{-1}$
s$^{-1}$.  A value $\tilde{f}_{V}\sim 10^{14}$ dyne cm$^{-1}$
is a conservative assumption (see Jones 1991) which,
with the estimated $\tilde{\sigma} > 10^{32}B_{12}^{-1}$
s$^{-1}$ for $N_{\mu} \ll N_{e}$, shows that the condition
is satisfied. However,
much higher spin-down rates exist during the propeller
phase of binary systems.  The model of Urpin, Geppert \&
Konenkov (1997) gives
\begin{eqnarray}
\dot{\Omega} = - (GM)^{3/7}R^{6/7}B^{2/7}\dot{M}^{6/7}I^{-1},
\end{eqnarray}
where $M$, $R$ and $I$ are, respectively, the neutron star
mass, radius and moment of inertia.  For typical values
($M = 1.4 M_{\odot}$, $R = 1.2\times 10^{6}$ cm,
$I = 10^{45}$ g cm$^{2}$ and an accretion rate on to the
Alfv\'{e}n surface of $\dot{M} = 10^{-10}M_{\odot}$
yr$^{-1}$) the inequality (24) gives,
\begin{eqnarray}
\tilde{f}_{V}\tilde{\sigma} > 3\times 10^{46}
B_{12}^{2/7}\Omega^{-2},
\end{eqnarray}
that is, $\tilde{f}_{V}\tilde{\sigma} > 8\times 10^{48}$
dyne cm$^{-1}$ s$^{-1}$ for the rotation periods of
$10^{2}$ s that are observed at the end of the spin-down
phase of binary evolution (see the review of
Verbunt \& van den Heuvel 1995).  Satisfaction of this
condition is much more problematic.

Previous papers concerned with the flux transport velocity
have reached differing conclusions.  In their seminal paper
on type II superconductivity in neutron stars, Baym, Pethick
\& Pines assumed that the viscous force on unit length of
moving proton vortex would arise from a dissipation rate
of the order of
\begin{eqnarray}
\pi\xi^{2}\sigma\tilde{E}^{2} \sim
\phi_{0}H_{c2}\sigma \frac{v_{L}^{2}}{c^{2}},
\end{eqnarray}
analogous with the theory of Bardeen \& Stephen (1965).
In equilibrium with the force per unit length
$f_{B}\phi_{0}/B$ derived from equation (20), it
would give extremely small velocities
($v_{L}\sim 10^{-16}$ cm s$^{-1}$).
In this expression, $\tilde{E}$ is the microscopic induction
field, $\sigma$ is the normal-state conductivity,
and $H_{c2}$ the upper critical field of the superconductor.
It assumes, in a laboratory superconductor, relaxation of the
vortex-core electron distribution function to isotropy in the
frame of the ion lattice.  But no equivalent interaction exists
in a proton superconductor and the theory is therefore
inapplicable (Jones 1987, 1991: Harrison 1991).

Muslimov \& Tsygan (1985) considered $v_{L}$ to be determined
by the steady-state equilibrium between a buoyancy force (the
derivative of the isotropic part of equation 18) and a
viscous force derived from magnetic scattering of electrons
by an isolated moving vortex (see also Harvey, Ruderman \&
Shaham 1986).  These calculations gave $v_{L}\sim 10^{-10}
- 10^{-8}$ cm s$^{-1}$ but were subject to the objection
that the viscous force used was not correct for the
problem concerned, which was the motion of the whole
triangular lattice, with fairly small intervortex spacing
$d = 4.9 \times 10^{-10}B_{12}^{-1/2}$ cm, through the
electron gas (Jones 1987, 1991; Harrison 1991).  These
authors considered the coupling of the lattice with
the electrons and recognized that, given the limited
interactions occurring in a superfluid system
well below the critical temperature, the electron
distribution function must relax to isotropy in the rest
frame of the lattice.  Jones (1987) neglected the supercurrent
screening condition (equation 15) and is therefore seriously
in error, as is Jones (1991) who obtained a result equivalent
to equation (10) but then incorrectly identified
the quantity
${\bf J}^{e}\cdot\delta\mbox{\boldmath$\mathcal{E}$}$
as the rate of dissipation.  The final conclusions of both
these papers must therefore be disregarded.  Harrison (1991)
considered the gravity-induced electric field (the term
$\mbox{\boldmath$\mathcal{E}$}_{0}$ of equation 3) which
must be present in both normal and superconducting systems.
His treatment of the electron-lattice interaction is based
on assumptions similar to those of Jones (1991).  It includes
a Lorentz force on the lattice, but differs from Jones
(1987, 1991) and the present paper in not considering
the proton superfluid Magnus force produced by the
difference ${\bf v}_{p0} - {\bf v}_{L}$.  This is an
effect arising from classical hydrodynamics (see Nozi\`{e}res
and Vinen 1966, Jones 1991) and there would appear to be no
doubt of its existence.  Harrison also observed that the
buoyancy force must be included in the equation for hydrodynamic
equilibrium, as in equation (19).
His final conclusion
is that $v_{L}\ll 10^{-16}$ cm s$^{-1}$.

The paper by Goldreich \& Reisenegger (1992) is concerned
primarily with matter whose composition is defined as being
chemically homogeneous (neutrons and charged components limited to
electrons and protons), without superfluidity. It introduces to
neutron star physics the concept of ambipolar diffusion (see also
Haensel, Urpin \& Yakovlev 1990).
This is movement of the charged components and magnetic flux
relative to the neutral part of the system.  It would also be
of relevance to the intermediate state in proton type I
superconductivity if that were present in the neutron star
interior. These authors
observe that, in analyzing such motion, it is essential
to divide particle flux vectors $N_{i}{\bf v}_{i}$ into
solenoidal and irrotational components.  Then solenoidal
motion, for chemically homogeneous systems, is limited only
by a viscous force. Its velocity is
$v^{s}_{ambip}\propto \tau_{pn}$, where $\tau_{pn}$ is derived
from $p-n$ scattering and is, in the
normal system, the principal collision relaxation time.
(The distinction is that solenoidal components do not change
local number densities, whereas irrotational components
produce chemical potential gradients and an imbalance
$\delta\mu = \delta\mu_{p} + \delta\mu_{e} - \delta\mu_{n}$,
which is the basis for the relationship
between stability and stratification emphasized by these
authors.)

The structure of the liquid core may be more complex than we
have assumed in the present paper.  Even if type II superconductivity
were present in the outer region, protons in the inner core might be of
type I.   Type II flux quanta $\phi_{0}$, on entering the inner region,
would merge to form an intermediate state of the type I superconductor
in which magnetic flux is confined to filaments of normal proton
Fermi liquid with macroscopic cross-sectional area.  The relaxation
time $\tau_{pn}$ derived for proton interaction with superfluid
neutrons (well below the critical temperature) is extremely long so
it might be thought that there would
be rapid ambipolar diffusion of these filaments under stresses
generated by flux movement in the outer (type II) region.  But this
is not necessarily so owing to the presence, in many equations of state,
of muons at inner-region matter densities.  The matter is then chemically
inhomogeneous in the sense considered by Goldreich \& Reisenegger.
Radial motion would produce a large chemical potential difference
$\delta\mu_{e} - \delta\mu_{\mu}$ because the muons are non-relativistic.
The consequence, following Goldreich \& Reisenegger, is that even
solenoidal particle fluxes are strongly inhibited by the stratification
and stability condition so that flux movement out of the type I region
would be slow, depending on weak-interaction transitions to remove
the imbalance.  The rates for these are strongly suppressed by the
proton (and possibly also the neutron) energy gap; also by the
requirement for energy-momentum
conservation in direct $\mu \rightleftharpoons e$ transitions on
the Fermi surfaces with neutrino-pair creation.

We emphasize that the flux velocity ${\bf v}_{L}$ given by equation
(22) can be of an order of magnitude different from the individual
drift velocities
${\bf v}_{e,\mu,p0}$.  In this respect, the motion differs from the
ambipolar diffusion described by Goldreich \& Reisenegger.  On the
other hand, there is a similarity between equation (22) and ambipolar
diffusion which is worth noting
in relation to the proton-vortex drift velocity obtained
by Ruderman, Zhu \& Chen (1998; equations 10 and 14 of that paper)
which is inversely proportional to an effective conductivity
$\sigma$.  It is defined by dissipation in an
induction field rather than by a viscous force.  The
explanation for this appears to be that, in
considering proton vortex motion, these
authors have not included both the Magnus force and
the screening condition given by equation (15) of the present
paper.  This appears to be the origin of the disagreement
with our drift velocity ${\bf v}_{L}$.  Also, the effective
conductivity of Ruderman, Zhu \& Chen is determined
principally by the component of the magnetic force on an
isolated moving proton vortex that is antiparallel with its
velocity, and for that reason is referred to as the magnetic
viscous or drag force.  The view of the present paper 
(see also Jones 1991)
is that this scattering is the process that produces the
polygonal electron orbits and, of course, is the source of
their irregularity.  The discussion which
immediately follows equation (3) expresses our view that
this process, represented in equations (1) and (3) by the
terms containing $\tilde{\bf B} - {\bf B}$, does not
contribute to the Boltzmann collision integral.

		 The work of this paper has been limited to density
regions containing only electrons, muons, protons and neutrons.
Most equations of state have hyperon thresholds that are likely
to be exceeded in a typical $1.4 M_{\odot}$ neutron star.
Thus a complete discussion of flux transport would need to
consider not merely the possibility of type I proton
superconductivity in the higher-density regions, but the
physics of systems with two or more superconducting
baryonic components.

\bsp

\label{lastpage}

\end{document}